\documentclass[12pt]{article}
\usepackage{authblk}
\usepackage{graphicx}
\usepackage{natbib} 
\usepackage{url} 
\usepackage{amsmath} 
\usepackage{amsfonts}
\usepackage{amssymb} 
\usepackage{bbm}
\usepackage{lineno}
\usepackage{setspace}
\usepackage{lscape} 
\usepackage{float} 
\usepackage{breakcites}
\usepackage{pdfpages}
\usepackage{xcolor}
\usepackage[colorlinks=true,citecolor=black,linkcolor=black,urlcolor=black,filecolor=black]{hyperref}
\usepackage{setspace}
\usepackage{booktabs}
\usepackage{makecell}
\usepackage{multirow}

\newcommand{\blind}{0}
\newcommand{\bs}{\boldsymbol}

\usepackage{ntheorem}
\makeatletter
\renewtheoremstyle{plain}
  {\item{\theorem@headerfont ##1\ ##2\theorem@separator}~}
  {\item{\theorem@headerfont ##1\ ##2\ (##3)\theorem@separator}~}
\makeatother

\makeatletter
\newcommand*{\addFileDependency}[1]{
  \typeout{(#1)}
  \@addtofilelist{#1}
  \IfFileExists{#1}{}{\typeout{No file #1.}}
}
\makeatother

\begin{document}

\def\spacingset#1{\renewcommand{\baselinestretch}%
{#1}\small\normalsize} \spacingset{1}


%
%

\if0\blind
{
 \title{\bf Spatial distribution and determinants of childhood vaccination refusal in the United States}
    \author[1]{Bokgyeong Kang}
    \author[2]{Sandra Goldlust}
    \author[3]{Elizabeth C. Lee}
    \author[4]{John Hughes}
    \author[5]{Shweta Bansal}
    \author[1]{Murali Haran}
    \affil[1]{Department of Statistics, Pennsylvania State University}
    \affil[2]{New York University School of Medicine}
    \affil[3]{Department of Epidemiology, Johns Hopkins Bloomberg School of Public Health}
    \affil[4]{College of Health, Lehigh University}
    \affil[5]{Department of Biology, Georgetown University}
    \date{}
  \maketitle
} \fi

\if1\blind
{
  \bigskip
  \bigskip
  \bigskip
  \begin{center}
  {\LARGE\bf Spatial distribution and determinants of childhood vaccination refusal in the United States}
\end{center}
  \medskip
} \fi

\begin{abstract}
Parental refusal and delay of childhood vaccination has increased in recent years in the United States. This phenomenon challenges maintenance of herd immunity and increases the risk of outbreaks of vaccine-preventable diseases. We examine US county-level vaccine refusal for patients under five years of age collected during the period 2012--2015 from an administrative healthcare dataset. We model these data with a Bayesian zero-inflated negative binomial regression model to capture social and political processes that are associated with vaccine refusal, as well as factors that affect our measurement of vaccine refusal. Our work highlights fine-scale socio-demographic characteristics associated with vaccine refusal nationally, finds that spatial clustering in refusal can be explained by such factors, and has the potential to aid in the development of targeted public health strategies for optimizing vaccine uptake.
\end{abstract}

\noindent%
\vfill

\newpage
\spacingset{1.5} 



\section{Introduction}
\label{sec:intro}

In recent years, vaccine hesitancy (i.e., a desire to delay or refuse vaccination, despite the availability of vaccination services) has resurged in the United States, challenging the maintenance of herd immunity for childhood infections \cite{MacDonald2015}. State-mandated school entry immunization requirements in the United States play an important role in achieving high coverage for childhood vaccines, but variations in vaccine exemption policies result in a patchwork of vaccine coverage across the country. A study based on the Centers for Disease Control's (CDC) National Immunization Survey (NIS) found that 14\% of parents with children aged 24--35 months refused or delayed one or more doses of vaccine of their children in 2009 \cite{smith2011parental}. A recent study showed that these trends have continued with an increase in the proportion of children with exemptions from school vaccination requirements in many states \cite{Zipfel2020}. Due to disruptions caused by the COVID-19 pandemic, routine childhood vaccination rates have reduced in the United States and globally, making any amount of vaccine refusal particularly dangerous \cite{gaythorpe2021impact,desilva2022association}.
The phenomenon of vaccine refusal has been shown to be associated with an increased risk for vaccine-preventable childhood diseases \cite[see][for a review]{Phadke2016}; as such, there is a large literature on monitoring child vaccination trends. However, much of this work relies on data from self-reported vaccination behavior or administrative records of exemptions from childhood vaccination laws and has been done on smaller spatial scales. What is urgently needed is surveillance of verified childhood vaccine refusal behavior at a fine geographic scale across the United States so that we may characterize trends in this behavior and identify pockets of susceptibility to childhood infections.

Previous studies have investigated the demographics of intentionally unvaccinated children and have found them more likely to be Caucasian males of married parents, with greater rates of college education, private school attendance, household income, and household size \cite{Smith2004,Gust2008,Birnbaum2013}. Parents engaging in vaccine hesitancy cite personal and religious beliefs as motivating their behavior, including concerns about vaccine safety, doubts regarding the necessity of immunizations for child health, and opposition to school immunization requirements \cite{Salmon2015,Omer2009,Salmon2009,Larson2014}. We note that determinants of vaccine refusal may differ from those of underimmunization, which is driven by vaccine unavailability or costs ineligibility.

In the United States, surveillance of vaccine uptake for childhood infections is limited in scope and spatial resolution. Much past work has relied on the CDC's NIS data or school vaccination exemption records to assess vaccine uptake across the United States. The NIS's samples are representative and comparable across states, but the survey suffers from low response rates and low spatial resolution (state-level) \cite{Smith2004,Salmon2006,Frieden2014,Hill2015}. By contrast, school vaccination exemption records have high response rates and can offer finer spatial resolution and vaccine-specific information. However, only a small number of states provide county-level exemption data, with varying data collection methods and data quality \cite{Zipfel2020}; additionally, school vaccination exemption records have been shown to have limited accuracy (e.g., one study found that 22\% of exempt children were in fact fully vaccinated \cite{Salmon2005}). Surveys have been conducted to investigate the determinants of parents' reluctance to vaccinate their children, but those studies are typically limited in scale and do not sufficiently  explore factors associated with immunization behavior \cite{Salmon2004survey,Salmon2005survey}. The lack of verified tools for assessing and quantifying vaccine underutilization prohibits the development of geographically-targeted and context-specific interventions \cite{Larson2015,Eskola2015}. By providing large-volume and high-resolution data, electronic health records offer new opportunities to study infectious diseases and related behaviors with greater accuracy and validation \cite{Lee2016,Bansal2016,Lee2018}. A limited number of studies have considered the use of ICD-9 codes for assessing vaccination status using claims data from the CDC’s Vaccine Safety Datalink (VSD) \cite{McCarthy2013,Glanz2013} and Kaiser Permanente managed care organization \cite{Lieu2015,Ackerson2021,Weinmann2020}; however, these studies have been limited to a small number of states.

Vaccine hesitancy behavior has been shown to be spatially clustered. Identifying this geographic clustering in vaccine hesitancy is important for controlling and eliminating vaccine-preventable diseases, as they can increase the risk of disease outbreaks and transmission \cite{Omer2008,Atwell2013,Ernst2012,Truelove2019,Masters2020,Gromis2022}. The spatial clustering patterns we identify will point to two alternative explanations for spatially dependent behavior: \textit{social influence} or \textit{social selection} \cite{lucila2022}. Spatial dependence may be produced by social influence or diffusion of vaccination behavior between neighboring areas through learning of social norms and practices; this makes understanding social interactions between interdependent units crucial to understanding behavior. Alternatively, neighboring areas may experience social selection and independently adopt similar behaviors because they share characteristics that promote the behavior; this means that the clustering in vaccination behavior only reflects geographic clustering in underlying drivers. While these two processes are not mutually exclusive, learning the determinants of spatial dependence is critical to dynamic modeling of behavior as well as to informing policy \cite{lucila2022}. However, due to lack of empirical evidence, past disease-behavior dynamical models have assumed the former alternative and modeled vaccination behavior as a contagion process \cite{Bauch2005,Oraby2014}. Large-scale data on refusal and spatial models can provide an opportunity to study this further.

In this article, we use large-scale and high-resolution data of vaccine refusal across the United States to estimate local
pockets of vaccine refusal across the United States for the years 2012-2015. We also conduct an ecological analysis of the socio-economic determinants of vaccine refusal and identify clusters of low vaccine protection. A reliable spatial and temporal understanding of vaccine underutilization, based on both behavioral data and an understanding of the underlying drivers of behavior, could play a critical role in clinical practice and public health decision-making.

\section{Material and methods}
\label{sec:method}

\setlength{\tabcolsep}{2pt}
\begin{table}[!b]
    \centering
    \small
    \caption{Description of the covariate variables}
    \label{tab:predictorsinmodel}
    \begin{tabular}{lllll}
        \toprule
        \multicolumn{2}{l}{Variable} & Description & Year(s) & Source\\ \midrule
        \multicolumn{3}{l}{\textit{Measurement variables}}\\\cmidrule{2-5}
         & \makecell[l]{Physician-patient\\ interactions} & \makecell[l]{Number of physician-patient\\ interactions} & 2012--2015 & IMS Health\\\cmidrule{2-5}
         & Health insurance & \makecell[l]{Proportion of people with health\\ insurance} & 2011--2014 & SAHIE\\\cmidrule{2-5}
         & Pediatrician reporting & \makecell[l]{Rate at which a pediatrician\\ voluntarily reports non-billable\\ diagnoses} & 2012--2015 & \makecell[l]{CDC Natality;\\IMS Health} \\ \midrule
        \multicolumn{3}{l}{\textit{Demographic or socioeconomic variables}}\\\cmidrule{2-5}
         & Household size & \makecell[l]{Average number of individuals\\ living in a single household} & 2011--2015 & ACS\\\cmidrule{2-5}
        & \makecell[l]{Religious\\ congregations} & \makecell[l]{Per capita number of congregations\\ of religions historically opposed to\\ vaccination} & 2010 & RCMS\\\cmidrule{2-5}
         & \makecell[l]{Limited English\\ proficiency} & \makecell[l]{Proportion of people who are not\\ proficient in English} & 2011--2015 & ACS \\\cmidrule{2-5}
         & Private school & \makecell[l]{Proportion of children who attend\\ private school} & 2011--2015 & ACS\\\cmidrule{2-5}
         & High income & \makecell[l]{Proportion of people in the upper\\ 20\% quantile of income in the US} & 2010--2014 & ACS\\\cmidrule{2-5}
         & Same area & \makecell[l]{Proportion of people living in\\ the same county one year prior} & 2010--2014 & AHRF\\\cmidrule{2-5}
         & State law leniency & \makecell[l]{Exemption law effectiveness index} & \multicolumn{2}{l}{\cite{bradford2015some}} \\\cmidrule{2-5}
         & State autism & \makecell[l]{Among families with more than\\ 1 child, the proportion with a\\ current or past diagnosis of autism} & 2012 & NSCH
         \\
        \bottomrule
    \end{tabular}
    
    \medskip
    {\raggedright ACS = American Community Survey, AHRF = Area Health Resource Files, CDC = Centers for Disease Control and Prevention, NSCH = National Survey of Children's Health, RCMS = Religious Congregations and Membership Study, SAHIE = Small Area Health Insurance Estimate\par}
\end{table}

\subsection{Medical claims data}

Monthly reports for vaccine refusal among patients under five years of age were obtained from a database of U.S. medical claims managed by IMS Health (a subset of the U.S.-based Dx Database). Claims were submitted from both private and government insurance providers, and data were aggregated according to U.S. five-digit ZIP codes and summarized by year from 2012 to 2015. We allocated all ZIP code data to US counties using the address-weighted ZIP Code Crosswalk files provided by the U.S. Department of Housing and Urban Development (HUD) and United States Postal Service (USPS). 
The reported data cover 126,049 cases of vaccine refusal for children under five years of age across 2,470,638,941 physician-patient interactions for people of all ages during the period of 2012--2015.
Vaccine refusal was identified with the International Classification of Disease, Ninth Revision (ICD-9) code for ``vaccination not carried out'' (\texttt{V64}) with subcodes relating to non-medical reasons for refusal including caregiver refusal (\texttt{V64.5}, 83\%), patient refusal (\texttt{V64.6}, 15\%), and religious reasons (\texttt{V64.7}, 2\%).

\subsection{Predictors of vaccine hesitancy}

We conducted a literature review to identify hypothesized socio-economic determinants of vaccine hesitancy, such as income, education, household size, and religious group representation, and searched various publicly-available databases for county-level quantifiable measures of these potential drivers. Additionally, we used data from \cite{bradford2015some}, which quantifies the effectiveness of each state's vaccine exemption laws, and captured state-level data on the prevalence of autism in multi-child families to consider the finding of \cite{zerbo2018vaccination}. We selected predictor measures initially based on the quality and spatial resolution of available data. We also assessed pairwise correlation and variance inflation factors in order to minimize multi-collinearity before arriving at the final set of predictor measures (bottom of Table~\ref{tab:predictorsinmodel}). We find these socio-economic predictors to be relatively stable over the four-year period of study and thus represent them to be spatially varying (i.e., county scale) but not temporally varying. 
The summary statistics for these demographic or socioeconomic variables are found in Supplemental Table S2.

\subsection{Predictors of measurement processes}

In order to account for variability in measurement bias in our medical claims data, we identified four conditions, all of which would have to be met for a vaccine refusal to be captured in the database: (1) an individual seeks pediatric health care from a provider, (2) the individual is insured, (3) the provider uses the claims database, and (4) the provider reports the vaccine refusal. To capture these measurement processes, we include the following covariates: health insurance coverage, and the number of physician-patient interactions as a measure of database coverage. Additionally, we quantify the likelihood that physicians in a given county are likely to voluntarily report a non-billable health behavior by modeling reports of low birthweight by physicians (as reported while providing routine medical care during infancy) from our database against verified and complete data on birthweight from the National Center for Health Statistics Vital Statistics dataset in a linear model. Following exploratory analyses, which included assessing multicollinearity, we selected for inclusion in our model three measurement factors (top of Table~\ref{tab:predictorsinmodel}) representing these measurement mechanisms. We find reasonable year-to-year variability in the measurement factors over the four years of our study, and thus we represent them in a spatially-varying (i.e. county-scale) and temporally-varying (i.e. yearly) manner. The summary statistics of these measurement variables in each year are found in Supplemental Table S3.

The final dataset contains records for approximately 3,000 counties each year. All predictor data were centered and standardized for use in the model. 

\subsection{Model overview}
\label{subsec:model}

From a statistical standpoint, several important features of the data must be considered. First, the data are potentially zero-inflated as over 65\% of the counties reported no refusals in any given year. We assume that some of the zero observations were caused by imperfect detection of vaccine refusal cases through medical claims due to spatial variation in healthcare access and insurance rate. An appropriate model should account for imperfect detection. Second, given the wide range of positive refusal counts, the model for these data should accommodate potential overdispersion. We very briefly describe the main features of our proposed model. Details of inferential procedure and model choice can be found in the supplementary material. 

We consider a zero-inflated negative binomial model \cite{greene1994zinb,agresti2015foundations}, a common choice for modeling data with numerous zeros and overdispersion. This model consists of two components: a Bernoulli component and a negative binomial component. The Bernoulli component accounts for imperfect detection by classifying whether a zero is caused by imperfect detection. Then the entire dataset is divided into two groups: a missing data group with zeros caused by imperfect detection and a detected refusal group with the remaining zeros and positive observations. The zeros in the missing data group arise when some people refused vaccination but none of those refusals was captured in the database. A zero in the detected refusal group arose when no one refused vaccination in a given county and year. The negative binomial component models the observations in the detected refusal group and accounts for variability in the incidence of vaccine refusal under perfect detection.

Formally, in county $i$ and year $t$ we assume that the observed incidence $y_{it}$ of vaccine refusal is generated from the following model:
\begin{align}
    y_{it} \left\{ \begin{array}{ll}
        = 0 & \mbox{w.p. } 1 - \pi_{it}\\
        \sim \textup{NB}(\mu_{it}, \theta) & \mbox{w.p. } \pi_{it},
    \end{array} \right.
    \label{eq:zinb}
\end{align}
where $\pi_{it}$ is the probability of $y_{it}$ being in the detected group, $\mu_{it}$ represents the mean number of refusal cases under perfect detection, and $\theta$ is dispersion of counts. $\textup{NB}(a, b)$ represents the negative binomial distribution with mean $a$ and dispersion $b$. 
For detection probability $\pi_{it}$ and mean incidence under perfect detection $\mu_{it}$ we assume the logistic linear regression model and log-linear regression model, respectively:
\begin{align}
        \textup{logit}(\pi_{it}) &= \bs{x}_{it}^\top \bs{\beta}_1 \label{eq:logitpi}\\
        \log(\mu_{it}) &= \bs{x}_{it}^\top \bs{\beta}_2 \label{eq:logmu}
\end{align}
where $\bs{x}_{it}$ is a vector of predictors for county $i$ in year $t$, and $\bs{\beta}_1$ and $\bs{\beta}_2$ are vectors of regression coefficients.

\subsubsection{Modeling spatial dependence}

To identify the source of geographic clustering in vaccine refusal, we consider spatial dependence in addition to the spatial structure provided by our predictors. For mean $\mu_{it}$ we assume spatial generalized linear mixed models \cite[SGLMMs;][]{Diggle1998} to account for the potential residual spatial dependence:
\begin{align}
    \log (\mu_{it}) &= \bs{x}_{it}^\top \bs{\beta}_2 + W_{i}, \label{eq:logmu_sglmm}
\end{align}
where $W_{i}$ are spatial random effects for mean incidence of refusal under perfect detection. For computational efficiency and interpretability, we use a Bayesian spatial filtering version \cite{hughes2017spatial} of the SGLMM above. Details of the model and inferential procedure can be found in the supplementary material. 

We assume that the fixed effects ($\bs{x}_{it}^\top \bs{\beta}_2$) can account for spatial dependence caused by social selection, whereas the spatial random effects ($W_{i}$) may suggest spatial dependence caused by social influence in addition to social selection. Our null hypothesis is that the model \eqref{eq:logmu} is the best model for our data, which means that there is no additional effect of social influence beyond associations captured by plausible social selection factors. The alternative hypothesis is that the model \eqref{eq:logmu_sglmm} provides a better fit. We will compare two models using Bayesian model choice criteria.

\section{Results}
\label{sec:results}

We present the results from our two-level spatial hierarchical Bayesian model to estimate the geographic distribution of vaccine refusal informed by our high-coverage U.S. claims data on cases of vaccine refusal, as identified by healthcare providers during patient visits. Our model proceeds in two levels: 
first, we account for imperfect detection of vaccine refusal cases caused by spatial variability in measurement bias in our medical claims data.
Then, based on our spatial estimates of probabilities of detection, we can infer the abundance and geographic distribution of cases. The model thus allows us to carry out surveillance of vaccine refusal behavior across the United States. In addition, our model allows us to infer the social, economic, and health policy factors associated with vaccine refusal.

\subsection{Model fit and validation}

\begin{figure}[!b]
    \centering
    \includegraphics[width = \textwidth]{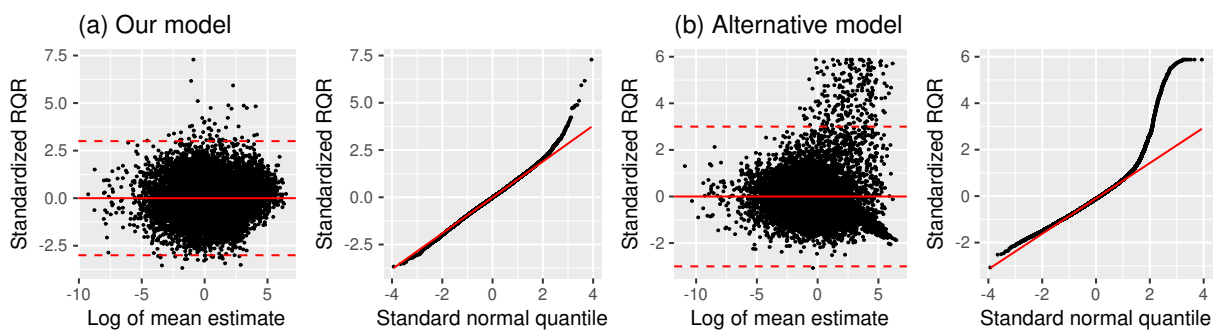}
    \caption{(a) Residual analysis plots from our model which can accommodate over-dispersion. (b) Residual analysis plots from an alternative model which assumes equi-dispersion. Our model fits the data much better than the alternative model.}
    \label{fig:rqr}
\end{figure}

Residual diagnosis is used for assessing the goodness-of-fit of regression models. However, for discrete data, standard residuals are distributed far from normality even though the model is well specified. Randomized quantile residuals (RQRs) are shown to provide low type I error and great statistical power for detecting model misspecification for count regression models including zero-inflated models \cite{Dunn1996,Feng2020}. In Figure~\ref{fig:rqr} we present residual analysis plots for the RQRs stemming from our model and an alternative model. The alternative model employs a Poisson distribution for describing counts under perfect detection and only assumes equi-dispersion (the mean of counts equals the variance) while our model uses a negative binomial distribution and can accommodate equi- and over-dispersion (the mean of counts equals the variance or below). The alternative model provided infinite values of RQR which were excluded in the plots. Figure~\ref{fig:rqr} shows that our model fits the data much better than the alternative model. This suggests the need for models that allow for overdispersion in the counts.

 \begin{figure}
    \centering
    \includegraphics[width = 0.8\textwidth]{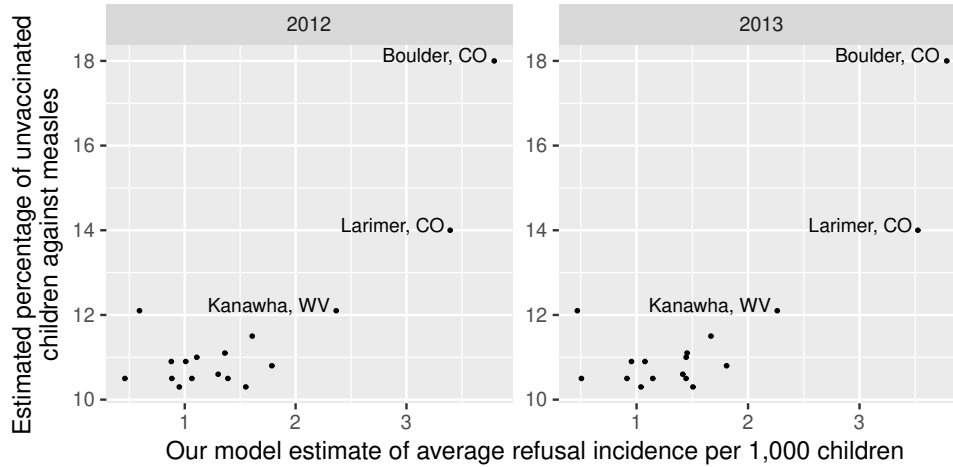}
    \caption{Our model estimate of average refusal incidence per 1,000 children under perfect detection compared to county-level estimated percentage of unvaccinated children against measles \cite{Salmon2015} by year.}
    \label{fig:salmon}
\end{figure}

To validate our model we compared our model estimates to county-level school vaccine exemption data \cite{Zipfel2020}. The school exemption data were collected at the county level from US states for which school vaccine exemption information was freely available from the Department of Health website. These data contain the total exemption rate, which represents the percentage of enrolled students with a vaccine exemption in any given county and year. We obtained model estimates of average refusal cases ($\mu_{it}$) per 1,000 children under perfect detection, for comparison with the total exemption rate. 
We merged our model estimates and the school exemption rate according to county and year, which provides 16 states to compare.
With a significance level of 0.05, we observed a strong association between our model estimates and the exemption rates in Florida, New Jersey, and Alabama; a moderate association in Oregon, Massachusetts, and Arizona; a weak association in California, Iowa, Washington, Minnesota, South Dakota, and Virginia; no association in Connecticut, New York, Pennsylvania, and Maine.
Corresponding results are found in Supplemental Table S4 and Figure S3.


\begin{figure}[!t]
    \centering
    \includegraphics[width=\textwidth]
    {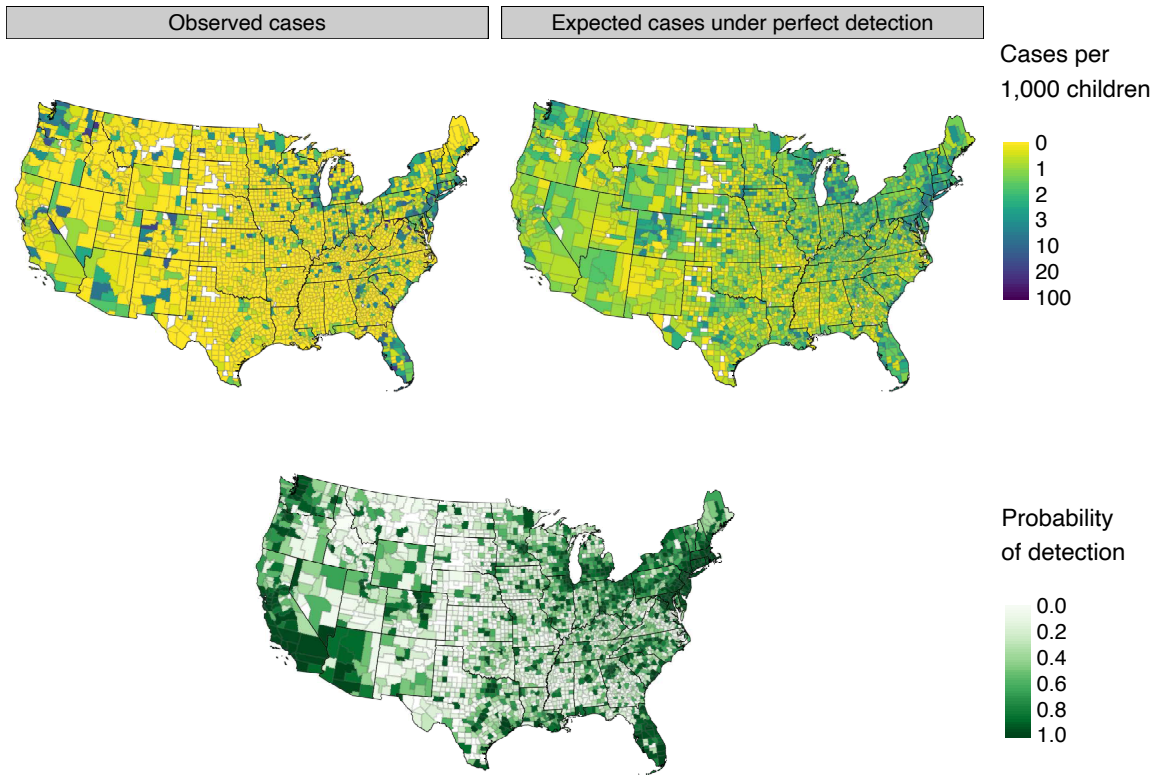}
    \caption{Top left: Observed refusal incidence per 1,000 children under five years of age in 2015. Top right: Estimated average refusal incidence per 1,000 children under perfect detection in 2015. Bottom: Model estimates of detection probability in 2015. Note: The reason for our choice of scale is that the data are zero-inflated and only a few counties have large counts.}
    \label{fig:main}
\end{figure}

We also compared our model estimates to county-level estimates of measles vaccination coverage \cite{Salmon2015}, in which the county-level percentage of unvaccinated children against measles were estimated using data from 2010-2013 NIS of households with 19- to 35-month-old children. \cite{Salmon2015} identified 20 counties with the highest values of estimated percentage of unvaccinated children. Figure~\ref{fig:salmon} shows our model estimates compared to \cite{Salmon2015}'s estimates for the overlapping counties and years. Both our study and \cite{Salmon2015}'s study found that Boulder County, Colorado, Larimer County, Colorado, and Kanawha County, West Virginia have the highest rates of vaccine refusal.

\subsection{Estimated spatial distribution of refusal cases}
\label{subsec:estimate}

Based on our model estimates, we find substantial heterogeneity in vaccine refusal and the probability of detection of refusal in our data both between and within states.
In Figure~\ref{fig:main} (top) we show observed cases of vaccine refusal for 2015 compared with model estimates of average refusal cases ($\mu_{it}$) under perfect detection. At the bottom of Figure~\ref{fig:main}, we show model estimates of detection probability in 2015. Corresponding model estimates for 2012--2014 are found in Supplemental Figures S4 and S5.

\subsection{Predictors of the detection of vaccine refusal}
\label{subsec:sigpred}

\begin{figure}[!t]
    \centering
    \includegraphics[width=\textwidth]{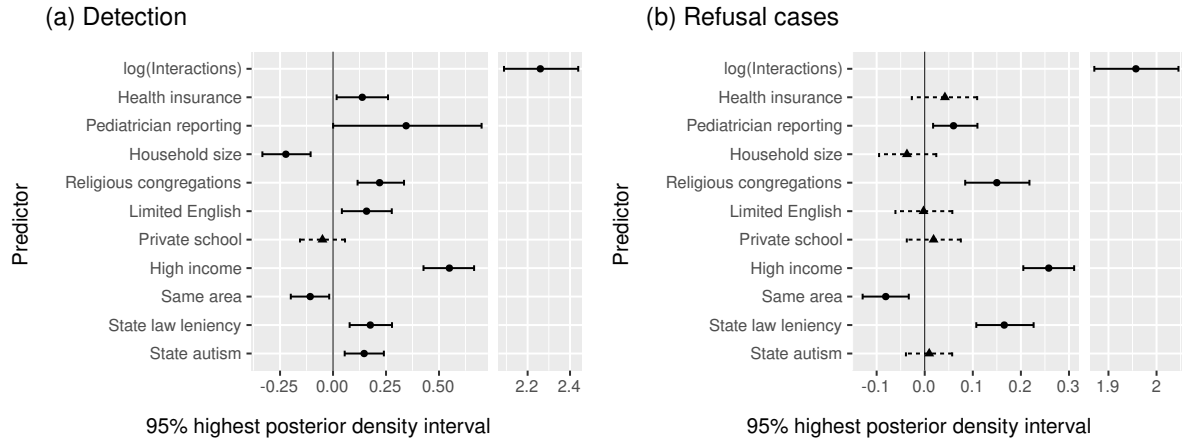}
    \caption{Estimated posterior means (shaded dots or triangles) and 95\% highest posterior density (HPD) intervals (horizontal solid or dashed bars) for predictor coefficients. The shaded dot and horizontal solid bar represent that the HPD interval does not include zero. The triangle and horizontal dashed bar represent that the HPD interval includes zero.}
    \label{fig:coef}
\end{figure}

Our proposed model allows us to identify determinants of the measurement of a vaccine refusal case. 
The probability of an observation being in the detected refusal group was modeled as the logistic linear regression model as in \eqref{eq:logitpi}.
Figure~\ref{fig:coef} (a) presents the estimate and 95\% credible interval for each regression coefficient for detection of vaccine refusal.
We found that the number of total physician-patient interactions (as a measure of database coverage in that location) are overwhelmingly predictive of refusal measurement. We also found that other healthcare-related measurements such as physician reporting may be predictive of refusal detection, but insurance coverage is less predictive. We also found that communities with high incomes, small household sizes, religions historically opposed to vaccination, limited English proficiency, lack of continuity of care, high incidence of autism, and high leniency in the state's vaccination laws are more likely to have detections of vaccine refusal.

\subsection{Predictors of the abundance of vaccine refusal}
With the second component of our model, we investigate the determinants of vaccine refusal cases, once detected. 
The average number of refusal cases under perfect detection was modeled as the log linear regression model as in \eqref{eq:logmu}.
Figure~\ref{fig:coef} (b) shows the estimate and 95\% credible interval for each regression coefficient for refusal cases under perfect detection. 
Given perfect detection, we found that communities with high refusal are likely to be high-income, have increased access to care, have more physician reporting, have more leniency in state vaccination laws, and contain groups historically opposed to vaccination. We also found that communities with lack of continuity of care are more likely to refuse to vaccinate their children. 

\subsection{Spatial dependence in vaccine refusal}
\label{subsec:spatialeffect}

To characterize the structure and reveal the source of spatial dependence (social influence versus social selection) in vaccine refusal patterns, we consider spatial random effects in addition to the spatial structure provided by our predictors. Our conjecture is that fixed effects can account for spatial dependence caused by social selection while the spatial random effects may suggest spatial dependence produced by social influence in addition to social selection, among other explanations. We compared the model with fixed effects only versus the model with fixed and spatial random effects via the widely applicable Watanabe--Akaike information criterion \cite[WAIC;][]{Watanabe2010}.
WAIC, like all information criteria, rewards out-of-sample predictive fit while penalizing model complexity. The model with fixed effects only gave WAIC = 32,838 with standard error = 461. The model with fixed and spatial random effects yielded WAIC = 32,535 with standard error = 454. Considering the standard errors, we can conclude that there is no significant difference between the two models. This suggests that there is no additional effect of social influence beyond associations captured by plausible social selection factors. 

\section{Discussion}
\label{sec:discussion}

We use a large-scale database of medical claims to create county-level Bayesian maps of the occurrence of vaccine refusal in the United States for the years 2012-2015. We also quantify the association between various socio-economic and policy factors and population-level. Our work is unique in conducting large-scale surveillance of childhood vaccine refusal across the United States and to investigate the associated behaviors with simultaneously high spatiotemporal resolution and coverage. Our study illustrates the potential of big data to help monitor vaccine intake and investigate the determinants of vaccine hesitancy in the United States.

We find that communities that have high incomes, have religions historically opposed to vaccination, have discontinuity of care, and live in states with permissive immunization laws are more likely to have high rates of refusal for childhood diseases, corroborating previous studies \cite{Luman2003,Smith2004,Omer2006,Kim2007,Gust2008,Wei2009,Atwell2013, Smith2005,Allred2007,Salmon2015}. On the other hand, we note that these associations may not generalize to non-childhood or non-endemic infections; in fact, recent work has highlighted that mistrust in the healthcare system among marginalized communities may play a bigger role in COVID-19 vaccine refusal than the associations we find.

We have also examined determinants of spatial clustering in vaccine refusal. We found evidence for geographic clusters of vaccine refusal being produced by social selection (in which communities sharing characteristics independently adopt similar behaviors, leading to spatial clustering in refusal) and less evidence for social influence (in which vaccine refusal behavior is contagious and diffuses through neighboring communities, generating refusal clusters). 
This suggests that investing in targeted outreach to high-refusal communities to understand their concerns, rather than focusing on preventing the general spread of misinformation, may have a more visible impact on vaccination rates.
We may target communities that are at risk due to their own traits or are surrounded by a socio-cultural environment with a high propensity towards hesitancy; conventional public health strategies to lessen vaccination hesitancy, such as healthcare provider training and community health outreach programs, can be conducted once the communities have been identified \cite{lucila2022}. 

We conclude with a few important caveats. Consideration must be taken to avoid the ecological fallacy when interpreting our results. We conduct statistical inference at the county level, and we, therefore, are not trying to infer individual factors but rather ecological ones. Previous investigations of disease distributions have demonstrated sensitivity of statistical inference to spatial scale. We also acknowledge that our medical claims data are not vaccine-specific. Thus, our data may capture vaccine under-utilization for seasonal influenza or other vaccines for which alternative processes drive under-utilization. However, we believe the impact of non-childhood immunizations on our results is limited because we only analyze medical claims submitted for children under five years of age. Lastly, our data are from before the COVID-19 pandemic and thus do not reflect any shifts in the childhood vaccine hesitancy landscape due to pandemic public health campaigns and healthcare disruptions.

\section*{Acknowledgements}
This work was supported by the National Institute of General Medical Sciences of the National Institutes of Health under Award Number R01GM123007 (JH, SB, MH). We thank David Bradford for generously sharing their data on state vaccination laws.

\bibliographystyle{apalike}
\bibliography{cas-refs}

\begin{thebibliography}{}

\bibitem[Ackerson et~al., 2021]{Ackerson2021}
Ackerson, B.~K., Sy, L.~S., Glenn, S.~C., Qian, L., Park, C.~H., Riewerts,
  R.~J., and Jacobsen, S.~J. (2021).
\newblock {Pediatric vaccination during the COVID-19 pandemic}.
\newblock {\em Pediatrics}, 148.

\bibitem[Agresti, 2015]{agresti2015foundations}
Agresti, A. (2015).
\newblock {\em {Foundations of linear and generalized linear models}}.
\newblock John Wiley \& Sons.

\bibitem[Allred et~al., 2007]{Allred2007}
Allred, N.~J., Wooten, K.~G., and Kong, Y. (2007).
\newblock {The association of health insurance and continuous primary care in
  the medical home on vaccination coverage for 19-to 35-month-old children}.
\newblock {\em Pediatrics}, 119:S4--S11.

\bibitem[Alvarez$-$Zuzek et~al., 2022]{lucila2022}
Alvarez$-$Zuzek, L.~G., Zipfel, C., and Bansal, S. (2022).
\newblock {Spatial clustering in vaccination hesitancy: the role of social
  influence and social selection}.
\newblock {\em PLoS Computational Biology}, 18(10):e1010437.

\bibitem[Atwell et~al., 2013]{Atwell2013}
Atwell, J.~E., Otterloo, J.~V., Zipprich, J., Winter, K., Harriman, K., Salmon,
  D.~A., Halsey, N.~A., and Omer, S.~B. (2013).
\newblock {Nonmedical vaccine exemptions and pertussis in California, 2010}.
\newblock {\em Pediatrics}, 132:624--630.

\bibitem[Bansal et~al., 2016]{Bansal2016}
Bansal, S., Chowell, G., Simonsen, L., Vespignani, A., and Viboud, C. (2016).
\newblock {Big data for infectious disease surveillance and modeling}.
\newblock {\em The Journal of Infectious Diseases}, 214:S375--S379.

\bibitem[Bauch, 2005]{Bauch2005}
Bauch, C.~T. (2005).
\newblock {Imitation dynamics predict vaccinating behaviour}.
\newblock {\em Proceedings of the Royal Society B: Biological Sciences},
  272:1669--1675.

\bibitem[Birnbaum et~al., 2013]{Birnbaum2013}
Birnbaum, M.~S., Jacobs, E.~T., Ralston-King, J., and Ernst, K.~C. (2013).
\newblock {Correlates of high vaccination exemption rates among kindergartens}.
\newblock {\em Vaccine}, 31:750--756.

\bibitem[Bradford and Mandich, 2015]{bradford2015some}
Bradford, W.~D. and Mandich, A. (2015).
\newblock {Some state vaccination laws contribute to greater exemption rates
  and disease outbreaks in the United States}.
\newblock {\em Health Affairs}, 34(8):1383--1390.

\bibitem[DeSilva et~al., 2022]{desilva2022association}
DeSilva, M.~B., Haapala, J., Vazquez-Benitez, G., Daley, M.~F., Nordin, J.~D.,
  Klein, N.~P., Henninger, M.~L., Williams, J.~T., Hambidge, S.~J., Jackson,
  M.~L., et~al. (2022).
\newblock Association of the covid-19 pandemic with routine childhood
  vaccination rates and proportion up to date with vaccinations across 8 us
  health systems in the vaccine safety datalink.
\newblock {\em JAMA pediatrics}, 176(1):68--77.

\bibitem[Diggle et~al., 1998]{Diggle1998}
Diggle, P.~J., Tawn, J.~A., and Moyeed, R.~A. (1998).
\newblock {Model-based geostatistics}.
\newblock {\em Journal of the Royal Statistical Society: Series C (Applied
  Statistics)}, 47:299--350.

\bibitem[Dunn and Smyth, 1996]{Dunn1996}
Dunn, P.~K. and Smyth, G.~K. (1996).
\newblock {Randomized quantile residuals}.
\newblock {\em Journal of Computational and Graphical Statistics}, 5:236--244.

\bibitem[Ernst and Jacobs, 2012]{Ernst2012}
Ernst, K.~C. and Jacobs, E.~T. (2012).
\newblock {Implications of philosophical and personal belief exemptions on
  re-emergence of vaccine-preventable disease: the role of spatial clustering
  in under-vaccination}.
\newblock {\em Human Vaccines and Immunotherapeutics}, 8:838--841.

\bibitem[Eskola et~al., 2015]{Eskola2015}
Eskola, J., Duclos, P., Schuster, M., MacDonald, N.~E., Liang, X., Chaudhuri,
  M., Dube, E., Gellin, B., Goldstein, S., Larson, H., Manzo, M.~L., Reingold,
  A., Tshering, K., Zhou, Y., Guirguis, S., and Hickler, B. (2015).
\newblock {How to deal with vaccine hesitancy?}
\newblock {\em Vaccine}, 33:4215--4217.

\bibitem[Feng et~al., 2020]{Feng2020}
Feng, C., Feng, C., Li, L., and Sadeghpour, A. (2020).
\newblock {A comparison of residual diagnosis tools for diagnosing regression
  models for count data}.
\newblock {\em BMC Medical Research Methodology}, 20:1--21.

\bibitem[Frieden et~al., 2014]{Frieden2014}
Frieden, T.~R., Jaffe, D. W.~H., Kent, C.~K., Leahy, M.~A., Martinroe, J.~C.,
  Spriggs, S.~R., Starr, T.~M., Doan, Q.~M., King, P.~H., Roper, W.~L., Hill,
  C., Boulton, C. L.~M., Arbor, A., Caine, M. A.~V., Fielding, I. E.~J., Jones,
  T.~F., Khabbaz, T. F.~R., Maki, G. G.~D., Quinlisk, W.~P., Moines, D.,
  Remington, I. L.~P., and Schaffner, W.~W. (2014).
\newblock {National, state, and selected local area vaccination coverage among
  children aged 19–35 months — United States, 2013}.
\newblock {\em Morbidity and Mortality Weekly Report}, 63:748.

\bibitem[Gaythorpe et~al., 2021]{gaythorpe2021impact}
Gaythorpe, K.~A., Abbas, K., Huber, J., Karachaliou, A., Thakkar, N., Woodruff,
  K., Li, X., Echeverria-Londono, S., Ferrari, M., Jackson, M.~L., et~al.
  (2021).
\newblock Impact of covid-19-related disruptions to measles, meningococcal a,
  and yellow fever vaccination in 10 countries.
\newblock {\em Elife}, 10:e67023.

\bibitem[Glanz et~al., 2013]{Glanz2013}
Glanz, J.~M., Newcomer, S.~R., Narwaney, K.~J., Hambidge, S.~J., Daley, M.~F.,
  Wagner, N.~M., McClure, D.~L., Xu, S., Rowhani-Rahbar, A., Lee, G.~M.,
  Nelson, J.~C., Donahue, J.~G., Naleway, A.~L., Nordin, J.~D., Lugg, M.~M.,
  and Weintraub, E.~S. (2013).
\newblock {A population-based cohort study of undervaccination in 8 managed
  care organizations across the United States}.
\newblock {\em JAMA Pediatrics}, 167:274--281.

\bibitem[Greene, 1994]{greene1994zinb}
Greene, W.~H. (1994).
\newblock {Accounting for excess zeros and sample selection in Poisson and
  negative binomial regression models}.
\newblock Technical report, New York University.

\bibitem[Gromis and Liu, 2022]{Gromis2022}
Gromis, A. and Liu, K.~Y. (2022).
\newblock {Spatial clustering of vaccine exemptions on the risk of a measles
  outbreak}.
\newblock {\em Pediatrics}, 149:e2021050971.

\bibitem[Gust et~al., 2008]{Gust2008}
Gust, D.~A., Darling, N., Kennedy, A., and Schwartz, B. (2008).
\newblock {Parents with doubts about vaccines: which vaccines and reasons why}.
\newblock {\em Pediatrics}, 122:718--725.

\bibitem[Hill et~al., 2015]{Hill2015}
Hill, H.~A., Elam-Evans, L.~D., Yankey, D., Singleton, J.~A., and Kolasa, M.
  (2015).
\newblock {National, state, and selected local area vaccination coverage among
  children aged 19–35 months — united states, 2014}.
\newblock {\em Morbidity and Mortality Weekly Report}, 64:889--896.

\bibitem[Hughes, 2017]{hughes2017spatial}
Hughes, J. (2017).
\newblock {Spatial regression and the Bayesian filter}.
\newblock {\em arXiv preprint arXiv:1706.04651}.

\bibitem[Kim et~al., 2007]{Kim2007}
Kim, S.~S., Frimpong, J.~A., Rivers, P.~A., and Kronenfeld, J.~J. (2007).
\newblock {Effects of maternal and provider characteristics on up-to-date
  immunization status of children aged 19 to 35 months}.
\newblock {\em American Journal of Public Health}, 97:259--266.

\bibitem[Larson et~al., 2014]{Larson2014}
Larson, H.~J., Jarrett, C., Eckersberger, E., Smith, D. M.~D., and Paterson, P.
  (2014).
\newblock {Understanding vaccine hesitancy around vaccines and vaccination from
  a global perspective: a systematic review of published literature,
  2007-2012}.
\newblock {\em Vaccine}, 32:2150--2159.

\bibitem[Larson et~al., 2015]{Larson2015}
Larson, H.~J., Jarrett, C., Schulz, W.~S., Chaudhuri, M., Zhou, Y., Dube, E.,
  Schuster, M., MacDonald, N.~E., Wilson, R., Eskola, J., Liang, X., Gellin,
  B., Goldstein, S., Larson, H., Manzo, M.~L., Reingold, A., Tshering, K.,
  Duclos, P., Guirguis, S., and Hickler, B. (2015).
\newblock {Measuring vaccine hesitancy: the development of a survey tool}.
\newblock {\em Vaccine}, 33:4165--4175.

\bibitem[Lee et~al., 2018]{Lee2018}
Lee, E.~C., Arab, A., Goldlust, S.~M., Viboud, C., Grenfell, B.~T., and Bansal,
  S. (2018).
\newblock {Deploying digital health data to optimize influenza surveillance at
  national and local scales}.
\newblock {\em PLOS Computational Biology}, 14:e1006020.

\bibitem[Lee et~al., 2016]{Lee2016}
Lee, E.~C., Asher, J.~M., Goldlust, S., Kraemer, J.~D., Lawson, A.~B., and
  Bansal, S. (2016).
\newblock {Mind the scales: harnessing spatial big data for infectious disease
  surveillance and inference}.
\newblock {\em The Journal of Infectious Diseases}, 214:S409--S413.

\bibitem[Lieu et~al., 2015]{Lieu2015}
Lieu, T.~A., Ray, G.~T., Klein, N.~P., Chung, C., and Kulldorff, M. (2015).
\newblock {Geographic clusters in underimmunization and vaccine refusal}.
\newblock {\em Pediatrics}, 135:280--289.

\bibitem[Luman et~al., 2003]{Luman2003}
Luman, E.~T., McCauley, M.~M., Shefer, A., and Chu, S.~Y. (2003).
\newblock {Maternal characteristics associated with vaccination of young
  children}.
\newblock {\em Pediatrics}, 111:1215--1218.

\bibitem[MacDonald et~al., 2015]{MacDonald2015}
MacDonald, N.~E., Eskola, J., Liang, X., Chaudhuri, M., Dube, E., Gellin, B.,
  Goldstein, S., Larson, H., Manzo, M.~L., Reingold, A., Tshering, K., Zhou,
  Y., Duclos, P., Guirguis, S., Hickler, B., and Schuster, M. (2015).
\newblock {Vaccine hesitancy: definition, scope and determinants}.
\newblock {\em Vaccine}, 33:4161--4164.

\bibitem[Masters et~al., 2020]{Masters2020}
Masters, N.~B., Eisenberg, M.~C., Delamater, P.~L., Kay, M., Boulton, M.~L.,
  and Zelner, J. (2020).
\newblock {Fine-scale spatial clustering of measles nonvaccination that
  increases outbreak potential is obscured by aggregated reporting data}.
\newblock {\em Proceedings of the National Academy of Sciences of the United
  States of America}, 117:28506--28514.

\bibitem[McCarthy et~al., 2013]{McCarthy2013}
McCarthy, N.~L., Irving, S., Donahue, J.~G., Weintraub, E., Gee, J., Belongia,
  E., and Baggs, J. (2013).
\newblock {Vaccination coverage levels among children enrolled in the Vaccine
  Safety Datalink}.
\newblock {\em Vaccine}, 31:5822--5826.

\bibitem[Omer et~al., 2008]{Omer2008}
Omer, S.~B., Enger, K.~S., Moulton, L.~H., Halsey, N.~A., Stokley, S., and
  Salmon, D.~A. (2008).
\newblock {Geographic clustering of nonmedical exemptions to school
  immunization requirements and associations with geographic clustering of
  pertussis}.
\newblock {\em American Journal of Epidemiology}, 168:1389--1396.

\bibitem[Omer et~al., 2006]{Omer2006}
Omer, S.~B., Pan, W.~K., Halsey, N.~A., Stokley, S., Moulton, L.~H., Navar,
  A.~M., Pierce, M., and Salmon, D.~A. (2006).
\newblock {Nonmedical exemptions to school immunization requirements: secular
  trends and association of state policies with pertussis incidence}.
\newblock {\em JAMA}, 296:1757--1763.

\bibitem[Omer et~al., 2009]{Omer2009}
Omer, S.~B., Salmon, D.~A., Orenstein, W.~A., deHart, M.~P., and Halsey, N.
  (2009).
\newblock {Vaccine refusal, mandatory immunization, and the risks of
  vaccine-preventable diseases}.
\newblock {\em New England Journal of Medicine}, 360:1981--1988.

\bibitem[Oraby et~al., 2014]{Oraby2014}
Oraby, T., Thampi, V., and Bauch, C.~T. (2014).
\newblock {The influence of social norms on the dynamics of vaccinating
  behaviour for paediatric infectious diseases}.
\newblock {\em Proceedings of the Royal Society B: Biological Sciences}, 281.

\bibitem[Phadke et~al., 2016]{Phadke2016}
Phadke, V.~K., Bednarczyk, R.~A., Salmon, D.~A., and Omer, S.~B. (2016).
\newblock {Association between vaccine refusal and vaccine-preventable diseases
  in the United States: a review of measles and pertussis}.
\newblock {\em JAMA}, 315:1149--1158.

\bibitem[Salmon et~al., 2015]{Salmon2015}
Salmon, D.~A., Dudley, M.~Z., Glanz, J.~M., and Omer, S.~B. (2015).
\newblock {Vaccine hesitancy: causes, consequences, and a call to action}.
\newblock {\em Vaccine}, 33:D66--D71.

\bibitem[Salmon et~al., 2004]{Salmon2004survey}
Salmon, D.~A., Moulton, L.~H., Omer, S.~B., Chace, L.~M., Klassen, A.,
  Talebian, P., and Halsey, N.~A. (2004).
\newblock {Knowledge, attitudes, and beliefs of school nurses and personnel and
  associations with nonmedical immunization exemptions}.
\newblock {\em Pediatrics}, 113:e552--e559.

\bibitem[Salmon et~al., 2005a]{Salmon2005}
Salmon, D.~A., Moulton, L.~H., Omer, S.~B., DeHart, M.~P., Stokley, S., and
  Halsey, N.~A. (2005a).
\newblock {Factors associated with refusal of childhood vaccines among parents
  of school-aged children: a case-control study}.
\newblock {\em Archives of Pediatrics \& Adolescent Medicine}, 159:470--476.

\bibitem[Salmon et~al., 2005b]{Salmon2005survey}
Salmon, D.~A., Omer, S.~B., Moulton, L.~H., Stokley, S., DeHart, M.~P., Lett,
  S., Norman, B., Teret, S., and Halsey, N.~A. (2005b).
\newblock {Exemptions to school immunization requirements: the role of
  school-level requirements, policies, and procedures}.
\newblock {\em American Journal of Public Health}, 95:436--440.

\bibitem[Salmon et~al., 2006]{Salmon2006}
Salmon, D.~A., Smith, P.~J., Navar, A.~M., Pan, W.~K., Omer, S.~B., Singleton,
  J.~A., and Halsey, N.~A. (2006).
\newblock {Measuring immunization coverage among preschool children: past,
  present, and future opportunities}.
\newblock {\em Epidemiologic Reviews}, 28:27--40.

\bibitem[Salmon et~al., 2009]{Salmon2009}
Salmon, D.~A., Sotir, M.~J., Pan, W.~K., Berg, J.~L., Omer, S.~B., Stokley, S.,
  Hopfensperger, D.~J., Davis, J.~P., and Halsey, N.~A. (2009).
\newblock {Parental vaccine refusal in Wisconsin: a case-control study}.
\newblock {\em Wisconsin Medical Journal (WMJ)}, 108:23.

\bibitem[Smith et~al., 2004]{Smith2004}
Smith, P.~J., Chu, S.~Y., and Barker, L.~E. (2004).
\newblock {Children who have received no vaccines: who are they and where do
  they live?}
\newblock {\em Pediatrics}, 114:187--195.

\bibitem[Smith et~al., 2011]{smith2011parental}
Smith, P.~J., Humiston, S.~G., Marcuse, E.~K., Zhao, Z., Dorell, C.~G., Howes,
  C., and Hibbs, B. (2011).
\newblock Parental delay or refusal of vaccine doses, childhood vaccination
  coverage at 24 months of age, and the health belief model.
\newblock {\em Public health reports}, 126(2\_suppl):135--146.

\bibitem[Smith et~al., 2005]{Smith2005}
Smith, P.~J., Santoli, J.~M., Chu, S.~Y., Ochoa, D.~Q., and Rodewald, L.~E.
  (2005).
\newblock {The association between having a medical home and vaccination
  coverage among children eligible for the vaccines for children program}.
\newblock {\em Pediatrics}, 116:130--139.

\bibitem[Truelove et~al., 2019]{Truelove2019}
Truelove, S.~A., Graham, M., Moss, W.~J., Metcalf, C. J.~E., Ferrari, M.~J.,
  and Lessler, J. (2019).
\newblock {Characterizing the impact of spatial clustering of susceptibility
  for measles elimination}.
\newblock {\em Vaccine}, 37:732--741.

\bibitem[Watanabe, 2010]{Watanabe2010}
Watanabe, S. (2010).
\newblock {Asymptotic equivalence of Bayes cross validation and widely
  applicable information criterion in singular learning theory}.
\newblock {\em Journal of Machine Learning Research}, 11:3571--3594.

\bibitem[Wei et~al., 2009]{Wei2009}
Wei, F., Mullooly, J.~P., Goodman, M., McCarty, M.~C., Hanson, A.~M., Crane,
  B., and Nordin, J.~D. (2009).
\newblock {Identification and characteristics of vaccine refusers}.
\newblock {\em BMC Pediatrics}, 9:1--9.

\bibitem[Weinmann et~al., 2020]{Weinmann2020}
Weinmann, S., Irving, S.~A., Koppolu, P., Naleway, A.~L., Belongia, E.~A.,
  Hambidge, S.~J., Jackson, M.~L., Klein, N.~P., Lewin, B., Liles, E., Marin,
  M., Smith, N., Weintraub, E., and Chun, C. (2020).
\newblock {Incidence of herpes zoster among varicella-vaccinated children, by
  number of vaccine doses and simultaneous administration of measles, mumps,
  and rubella vaccine}.
\newblock {\em Vaccine}, 38:5880--5884.

\bibitem[Zerbo et~al., 2018]{zerbo2018vaccination}
Zerbo, O., Modaressi, S., Goddard, K., Lewis, E., Fireman, B.~H., Daley, M.~F.,
  Irving, S.~A., Jackson, L.~A., Donahue, J.~G., Qian, L., et~al. (2018).
\newblock {Vaccination patterns in children after autism spectrum disorder
  diagnosis and in their younger siblings}.
\newblock {\em JAMA pediatrics}, 172(5):469--475.

\bibitem[Zipfel et~al., 2020]{Zipfel2020}
Zipfel, C.~M., Garnier, R., Kuney, M.~C., and Bansal, S. (2020).
\newblock {The landscape of childhood vaccine exemptions in the United States}.
\newblock {\em Scientific Data}, 7.

\end{thebibliography}

\end{document}